\documentclass{aa}  
\usepackage{natbib}    
\usepackage{graphicx}
\usepackage{txfonts}
\usepackage{lipsum}
\usepackage{subcaption}         
\usepackage{lscape}             
\usepackage{placeins}           
\usepackage{orcidlink}
\usepackage{xspace}
\usepackage{hyperref}

\newcommand{\vhone}{\texttt{VH-1}\xspace}
\newcommand{\mesa}{\texttt{MESA}\xspace}
\newcommand{\unsim}{\mathord{\sim}}

\begin{document}

   \title{3D Binary Evolution of a Merging Binary Black Hole Progenitor}

%
%
%

   \author{D.~Dickson\inst{\ref{affil:Leuven}}\corrauth{davey.dickson@kuleuven.be}        
        \and P. Marchant\inst{2}\email{Pablo.Marchant@UGent.be}
        \and JO. Sundqvist\inst{1}\email{jon.sundqvist@kuleuven.be}
        \and M. Fabry\inst{3}\email{matthias.fabry@villanova.edu}
        }

   \institute{Institute of Astronomy, KU Leuven,
Celestijnenlaan 200D, 3001 Leuven, Belgium \label{affil:Leuven}
\and Department of Physics and Astronomy,
Ghent University, Krijgslaan 299, 9000 Gent, Belgium
\and Villanova University, 800 Lancaster Ave, Villanova, PA 19085, USA}

   \date{Received Month DD, 2026}

 
  \abstract
   {Mass transfer is a highly dynamical 3D process that spans evolutionary timescales in binary stars. It is key to the production of gravitational waves through formation of binary black holes and other double compact objects.}
   {We integrate global 3D hydrodynamics into stellar evolutionary modeling to resolve mass transfer across nuclear timescales. We use this framework to resolve stable Case B mass transfer in a binary black hole progenitor and compare the resultant evolution with the predictions of 1D models.} 
   {We employ the shellular approximation in the stellar evolution code \mesa to map a 3D spatial structure by equipotential shells in the Roche geometry to maintain hydrostatic equilibrium across tidally-distorted surfaces. We employ the method of time-incremented multiscale evolution to alternate between stellar evolution and 3D hydrodynamics to span large timescales with dynamical feedback. For that purpose, we adapt the hydrodynamics code \vhone to generate a quasi-steady-state solution in 3D for each evolutionary timestep from the hydrostatic \mesa profile and further modify \mesa to defer to the mass transfer rates through L1 obtained by the corresponding 3D model. We employ this framework to examine a likely merging binary black hole progenitor system of a low-metallicity $30M_\odot$ donor star and $9.3M_\odot$ black hole from the end of the main sequence to core Carbon depletion.}
   {We modeled two complete mass transfer phases in 3D while self-consistently accounting for stellar evolution. These comprised a thermal-timescale early Case B phase and a nuclear-timescale late Case B phase. Our hybrid approach enabled this new regime of study by spanning $10^{5-6}\times$ longer timescales than previous 3D binary methods at comparable computational cost. The prescription of \cite{kolb&ritter} more closely predicted the results of 3D evolution than more recent mass loss prescriptions, except at the end of the first phase of Case B mass transfer; our 3D model remained well-contained by the outer Lagrange point, despite 1D predictions of substantial L3 overflow during this stage. L2 and L3 outflow streams formed even well below their respective filling factors at densities above the photospheric density of the donor. If taken into account, the L2 stream would account for $1.1\%$ of mass and $3.8\%$ of angular momentum lost during the early Case B transfer phase, and $0.3\%$ of mass and $1.0\%$ of angular momentum lost during late Case B, though some of that material is likely retained in a circumbinary disk. When present, the L3 flow carried specific angular momentum equivalent to an orbit at L3 while the L2 outflow carried $81.0\pm0.7\%$ of the specific angular momentum of an orbit at L2. We find incorporating 3D hydrodynamics during Case B mass transfer did not significantly impact expected merger times when compared with 1D predictions.}
   {We have constructed a new hybrid framework that robustly models 3D dynamics in tandem with stellar and orbital evolution on nuclear timescales to closely fit the evolution of interacting binaries. We found the majority of the simulated Case B evolution was well-approximated by the method of \cite{kolb&ritter}. Our model never overflowed L3 as predicted by 1D prescriptions. Dense L2/L3 outflows formed even while underflowing those equipotentials. Our model suggests these outflows can be neglected for the evolution of similar systems, but were dense enough to potentially impact X-ray observations.}


   \maketitle
\nolinenumbers

\section{Introduction}

Mass transfer (MT) is the process by which a donor star exchanges material with an accretor in a binary system \citep{lubow&shu}. MT phases may span millions or billions of years; during this time, they alter the evolutionary path of the binary and are in turn altered by that evolution (see \cite{marchant2024rev} for a recent review). Throughout this phase, multiple transport processes operate interdependently in multiple dimensions spanning the complete binary system, such as Roche lobe overflow (RLOF), wind RLOF, wind accretion, disk formation, contact formation, X-ray feedback, eccentricity changes, tidal interactions, or common envelope evolution \citep{paczynski1971rev}. These are not rare occurrences either; most massive stars undergo some or all of these binary interactions across their lives \citep{sana2012binary}.

Resolving even a subset of these multidimensional processes across the whole span of binary evolution has proven computationally prohibitive \citep{marchant2024rev,time}. Historically, MT models have been constrained to either span a brief moment in multiple dimensions (e.g. \citealp{bisikalo1998numerical,nazarenko2006,blondin2012hoyle,dickson2024,ryu2025binary}) or approximate the whole of the process by assuming spherical symmetry or even stricter analytical assumptions (e.g. \citealp{lubow&shu,kolb&ritter,tout1997rapid,langer2020OBbh,willcox2023popsynth}). Much of the recent work in simulating MT has been done with the Modules for Experiments in Stellar Astrophysics (\mesa; \citealp{mesa,mesa2,mesa3binary,mesa4,mesa5,mesa6}) code. In its standard distribution, \mesa performs detailed 1D binary evolution following the method of \cite{kolb&ritter} to approximate MT \citep{mesa3binary}. Recent works focused on MT and post-interaction binaries have sometimes added second-order terms to this same framework, as done by \cite{marchant2021roleofMT}, or considered different flow geometry, as done by \cite{cehula2023mtprescription}. The assumption of spherical symmetry in these models has also recently been revised to shift coordinates to increment the tidally-distorted gravitational potential rather than the radius \citep{fabry2022shell}. The key assumption of this method is the shellular approximation under which isobars are assumed to align identically with equipotential surfaces such that any given potential shell may be approximated as having a uniform density, pressure, and temperature \citep{vonzeipel1924radiative}.

Despite these advances, what happens to material lost by the donor in MT remains unclear; much of recent work assumes MT to be either conservative, rotationally-limited, or Eddington-limited. In conservative MT, all material lost by the donor is gained by the accretor. Under the rotationally-limited assumption, instead the accretor retains transferred material only up to the point of critical rotation. Eddington-limited MT alternatively assumes all transferred mass beyond the Eddington limit of the accretor to be ejected from the system with the specific AM of the accretor. Each of these approximations typically define all flow between components to occur through L1, with no loss from the system through L2/L3 outflow streams.

These assumptions are not without impact; the predicted formation rate of binary black holes (BBHs) may vary by $100\times$ or more between competing approximations \citep{impactonBBHs}, and the channel to form merging BBHs may span only a few $\%$ in mass ratio \citep{marchant2021roleofMT}. Gravitational waves, which can be detected during the mergers of BBHs and other double compact objects \citep{abbott2016}, provide an incomplete picture of key progenitor paths of BBHs which only binary evolution models can address \citep{gallegos2021bbh,marchant2021roleofMT}. Present research suggests the critical channel in forming BBHs may be stable MT, introducing multidimensional complexity across evolutionary time \cite{vanson2022stablemt,xu2026stablemt}.

Recent works have pushed 3D MT models to their limits, but even the most ambitious of these remain on dynamical timescales (hours$-$years) much shorter than the timescale of stable MT (kyrs$-$Gyrs) \citep{de2014}. These models have been used both to span phases of rapid evolution \citep{reichardt2019extending} and as a snapshot to characterize longer evolutionary stages \citep{ryu2025binary}. The recent time-incremented multiscale evolution (TIME) method enabled the first 3D examinations of MT across evolutionary timescales by using 3D models to capture the (quasi-)steady state\footnote{As MT removes mass from the donor star, no semidetached binary can be said to be truly in steady state. We refer to any quasi-steady state that persists across one or more orbits during thermal or nuclear-timescale MT simply as steady states to avoid further confusion.} and then evolve the binary analytically to initialize the next 3D model in the series \citep{time}. That work was limited to resolving evolutionary steps between models analytically without stellar response on thermal and nuclear timescales; while these can be safely neglected in most dynamical simulations, the stable MT channel depends critically on the thermal and nuclear evolution of the donor star \citep{hjellming1987thresholds,stabilitycriteria}.

In this work, we will combine 3D hydrodynamics with 1D binary evolution to examine stable MT in a merging BBH progenitor candidate. We detail our approach to 3D binary evolution in Section \ref{sec:meth}. We then apply this to model two Case B MT phases in 3D across $100~\text{kyr}$ nuclear timescales in Section \ref{sec:test}. Particularly, we track the mass and angular momentum (AM) loss in our test high-mass X-ray binary (HMXB) across both early and late Case B MT phases. We then validate the efficacy of this model and discuss our findings in Section \ref{sec:disc}. Finally, we summarize our conclusions in Section \ref{sec:conc}.

\section{Computational Methods} \label{sec:meth}

We construct a hybrid 3D+1D code to resolve the hydrodynamics of binary stars across evolutionary timescales. This relies on the TIME framework of \cite{time}, whereby we evolve the system in 1D as informed by the steady state of an associated 3D model for each 1D timestep. 

This process begins from a 1D binary evolution model (Section \ref{ssec:mesa}) which is then mapped into 3D space as a set of initial conditions (Section \ref{ssec:1Dto3D}). We evolve the mapped binary on a 3D hydrodynamic grid to achieve a steady state (Section \ref{ssec:vh1}). We employ this 3D steady state to inform the next step of 1D evolution and repeat the process for each subsequent step (Sections \ref{ssec:3Dto1D}). 

We implement this approach as a full hybrid application of \mesa for binary stellar evolution and Virginia Hydrodynamics One (\vhone; \citealp{vh1}) code for global 3D hydrodynamics. 

\subsection{1D Binary Evolution with \mesa} \label{ssec:mesa}

We employ the detailed stellar and binary evolution code \mesa for the 1D evolutionary steps in our hybrid model \citep{mesa3binary}. We make use of \mesa version \texttt{24.08.1} together with the \texttt{MESASDK} version \texttt{24.7.1}. \mesa uses an EOS constructed by blending in different regimes between OPAL \citep{Rogers2002}, FreeEOS \citep{Irwin2004}, HELM \citep{Timmes2000}, PC \citep{Potekhin2010}, Skye \citep{Jermyn2021} and SCVH \citep{Saumon1995}. Within the range covered by our simulations opacities are taken from OPAL \citep{Iglesias1993,
Iglesias1996} using relative metal fractions scaled from the solar values reported by \cite{Asplund+2009}. We make use of the \texttt{basic.net} nuclear network of MESA which includes $^1$H, $^3$He, $^4$He, $^{12}$C, $^{14}$N, $^{16}$O, $^{20}$Ne and $^{24}$Mg, with nuclear reaction rates taken from JINA REACLIB \citep{Cyburt2010} and NACRE \citep{Angulo1999}. Our simulations include metallicity scaled winds following the recipe of \cite{Brott+2011}, which combines the prescriptions of \cite{Nieuwenhuijzen_deJager1990}, \cite{Hamann+1995} and \cite{Vink+2001}. Convection is modeled using mixing length theory \citep{BohmVitense1958} as described by \cite{CoxGiuli1968} with a mixing length parameter $\alpha_\text{MLT}=2$. Convective boundaries are determined using the Ledoux criterion \citep{Ledoux1947}. We adopt $Z_\odot=0.0142$ for the solar metallicity following \cite{Asplund+2009}. Input files to reproduce the results of our \mesa simulations are made available at \href{https://zenodo.org/records/24324640}{\texttt{https://zenodo.org/records/24324640}}.

While our method supersedes the analytical MT prescriptions of \mesa with the results of the corresponding 3D model, we track both these computed MT rates at each evolutionary step for comparison.

To make this 1D profile compatible with full 3D hydrodynamical simulations without significant artifacts requires a close approximation of hydrostatic equilibrium at each timestep to provide stable initial conditions for 3D modeling. Following the method of \cite{fabry2022shell}, we computed a table of parameters with which to define a hydrostatic condition in the Roche geometry. This table spans 281 values of $\log q$ evenly ranging $-3$ to $3$, where $q$ is the mass ratio $q=M_a/M_d$ . For each value of $\log q$, the parameter table ranges across 751 radial values 0--$1.5~R_\text{RL}$. These radii represent the volume-equivalent radii of equipotential shells rather than spherical radii, and are equally distributed except for the range $0.93$--$1.0~R_\text{RL}$, which is given twice the grid density as elsewhere to better resolve the significant changes in geometry near Roche lobe filling. At each point in this grid, we compute the enclosed volume and surface area of the corresponding equipotential shell, as well as its specific moment of inertia and the equipotential-averaged effective gravity

\begin{equation}
    \langle g_{\rm eff}\rangle  =\langle\|\nabla\Psi\|\rangle\quad,
\end{equation}

where $\Psi$ is the effective gravitational potential in the corotating frame. These values are computed following the method of \cite{kippenhahn1970stellarstructure} and \cite{endal1976evolution} to adapt the spherically-symmetric stellar structure equations for the effect of 3D tidally-distorted geometry. By the shellular approximation, all intensive quantities are identical across an equipotential surface. We may therefore interpolate the stellar structure at any given point in 3D space from its effective gravitational potential in the Roche geometry. We utilize this aspherical geometry to enable the hydrostatic condition of \mesa evolution to account for the true forces applied to material within an overflowing star in 3D space \citep{fabry2022shell}. This approach has since been integrated into the most recent distribution, \texttt{MESA v26.04.1}.

By construction, this method maps a star to be hydrostatic in the corotating frame, which is not applicable to all stars in close binaries. Even in corotating stars, the shellular approximation does not fully hold without breaking thermal equilibrium \citep{vonzeipel1924radiative}. These competing limits particularly affect the stellar surface; as we expect the MT channel of our simulations to be driven by the deeper envelope, this is not expected to produce significant discrepancies in this particular application \citep{dickson2024}. Future models may expand this method to a wider range of binaries by an analogous computation with respect to a rigid rotating potential rather than the Roche geometry. Further work is also necessary to account for the aspherical thermal structure in 3D to compensate for the thermal and shellular structure mismatch in 3+1D simulations.

We may compute the deviation from true hydrostatic conditions in a given shell as

\begin{equation}
    \vec{a}=\frac{\nabla P}{-\rho}-\vec{g}_{\rm eff}\quad,
\end{equation}

where $\|\vec{a}\|=a=0$ describes a hydrostatic state. Then for a given shell, we obtain,

\begin{equation}
    \label{eqn:accel}
    a=\frac{\|\nabla P\|}{-\rho}-\langle g_{\rm eff}\rangle = -\left(\frac{1}{\rho}\frac{dP}{d \Psi} + 1\right) \langle g_{\rm eff}\rangle\quad.
\end{equation}

In 1D, $a=0$ is assumed, though values interpolated from the parameter table will differ and can be directly computed from the 1D profile. When mapping the profile onto 3D space (see Section \ref{ssec:1Dto3D} below), this deviation from true hydrostatic conditions manifests as artificial acceleration. This improved coordinate system reduces the artificial acceleration due to non-hydrostatic initialization in overflowing envelopes from as much as a few $\langle g_{\rm eff}\rangle$ when using a spherical-shell approach to a few percent of $\langle g_{\rm eff}\rangle$ even before refining the 1D mesh resolution (see Figure \ref{fig:1dmesh} and the accompanying discussion in Section \ref{ssec:1dres}). To improve our mapping onto 3D contours, we employ a spatial resolution of twice the \mesa standard \citep{mesa}. 

\begin{figure*}
    \centering
    \includegraphics[width=1\linewidth]{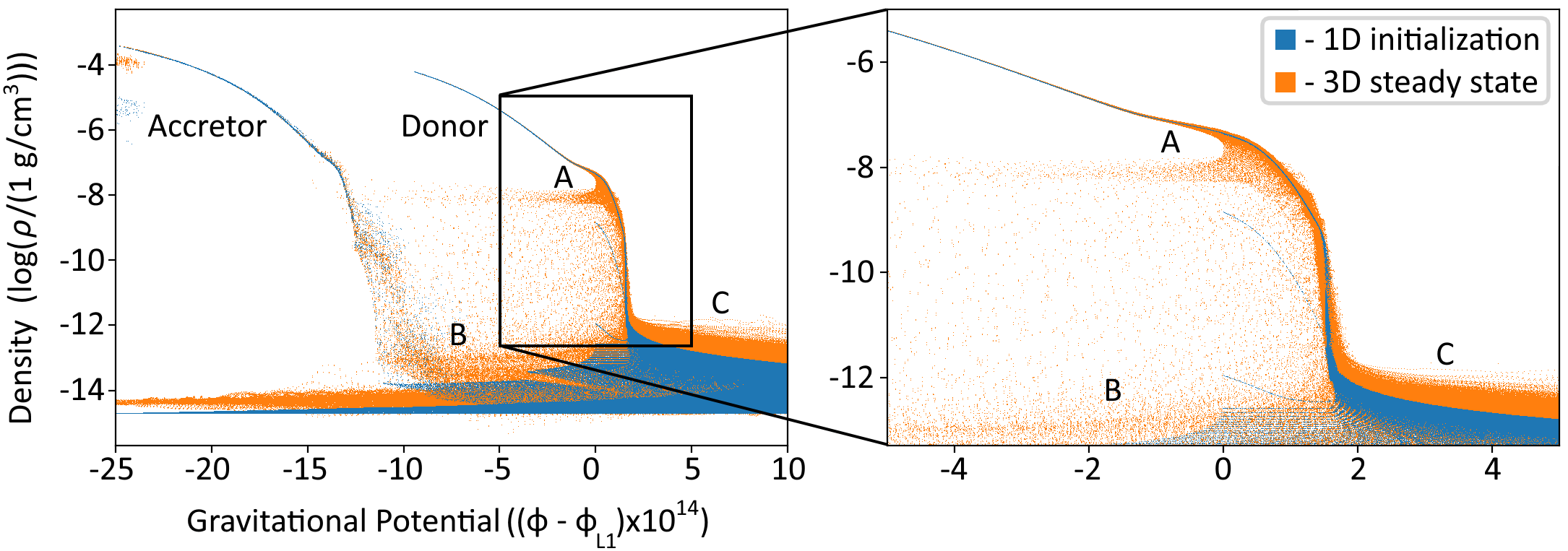}
    \caption{3D model relaxation of a barely overflowing Case A star+star binary of masses $24.1~M_\odot$ and $18.4~M_\odot$ from a hydrostatic \mesa profile. Each cell on a high resolution 3D \vhone grid after 10,000 steps of relaxation ($\unsim1~\text{day}$) is visualized as a pixel both in its initial state (blue), which constitutes the mapped \mesa profile plus a beta-law wind profile, and relaxed steady state (orange). The steady state differs primarily by the formation of the tidal stream (A) at $\phi=\phi_{L1}$, supplemental wind accretion (B), and enhancement of the donor wind (C). These effects and the dispersion visible in the 3D steady state arises primarily from induced flow dynamics not accounted for in the 1D profile.}
    \label{fig:relax}
\end{figure*}

\subsection{Initializing 3D from 1D} \label{ssec:1Dto3D}

With the 1D \mesa profile mapped by gravitational potential, we may initialize the corresponding 3D model. For each zone on our 3D grid, we compute the corresponding effective gravitational potential in the corotating frame following the Roche geometry. This enables us to initialize local variables by a logarithmic interpolation from the corresponding values at the nearest potentials in the \mesa profile. We import the full 1D profile into a set of initial conditions in 3D space, except that we truncate each initial 3D stellar envelope at the plane passing through L1 orthogonal to the line connecting the centers of masses in the system. This enables stars to overflow but prevents an overflowing star from initially filling both lobes in the 3D model. This approach differs from the truncation method employed to compute tidally-distorted geometry within \mesa, which relies on a splitting surface that passes through L1 locally orthogonal to the line connecting the centers of masses but curves to remain normal to the potential gradient \citep{fabry2022shell}. This discrepancy is not expected to meaningfully impact the relaxed state, which is not bound by either truncating surface.

In addition to specifying the initial state, we employ this equipotential map to define the at-rest radiation pressure everywhere within the stellar envelope(s). Rather than fixing the local radiation pressure, we employ the method of \cite{dickson2024} to mediate thermal domains. This approach maintains a radiative temperature gradient within slow-moving envelope material while enabling rapid flows to evolve adiabatically and stellar wind to evolve isothermally.

Initializing the wind does not demand the same level of detail as the stellar envelope(s), as stellar winds in \vhone models have been shown to converge on a timescale shorter than the relaxation of the tidal stream even with simplified initial conditions \citep{dickson2024}. We therefore initialize the wind with a simple beta law profile starting from the surface of the donor envelope, neglecting the wind of the initially less massive star. We import the donor's luminosity and effective temperature from \mesa to inform its dynamical evolution, but do not constrain the initial wind profile to precisely fit the wind speed or mass loss rate given by \mesa. The wind then launches dynamically from the envelope surface as previously in \cite{dickson2024}. 

An example of this model relaxation is given in Figure \ref{fig:relax} for a two-star system with the donor star exceeding its Roche lobe. The deep envelopes of both components remain quiet in the steady state, and we see the expected features emerge. The tidal stream launches in tandem with beamed wind through wind RLOF. The donor wind is enhanced as expected from similar models \citep{el2019wind,dickson2024}.

The above initial conditions are highly effective for two-star binaries, lowering the relaxation timescale of the model to be comparable to the transit timescale of the tidal stream. In binaries with compact object accretors, however, the formation and circularization timescales of the disk may be prohibitively long ($>10$ orbits at high filling factors $f=R_*/R_\text{RL}>1.2$). To combat this and abbreviate overall model relaxation, we initialize each 3D model with the disk of the prior model following the method of \cite{time}. 

We provide each model with one orbit for relaxation followed by three orbits to characterize the steady state.

\subsection{3D Hydrodynamics with \vhone} \label{ssec:vh1}

There are a number of codes capable of modeling binary systems in 3D, but application to 3D evolution is particularly demanding. This domain of study requires a highly efficient hydrodynamic code that can simultaneously resolve the entire domain of binary interaction.

To span evolutionary timescales with this method requires hundreds or thousands of dynamical models. Each dynamical model must relax to a quasi-steady state and then evolve for one or more orbits to provide adequate statistics to accurately replicate that steady state in 1D (see Section \ref{ssec:relax}). Under ideal conditions, this requires a model to span an orbit in $\lesssim10^3~\text{cpu-hrs}$ in order to complete a 3+1D TIME series within a month\footnote{This assumes 100 dynamical models with perfect scaling across 1024 processors, instantaneous relaxation to steady state, and a single orbit each for steady-state extraction.}. In practice, the limit is even more restrictive due to relaxation times and scaling inefficiencies. 

For each such orbit, the dynamical model must capture a significant fraction of the binary system. The donor envelope must be modeled to sufficient depth to capture the subsurface flows that feed L1 across a range of densities. This envelope must be able to interact with the donor wind and the area surrounding the accretor. That region may require modeling direct wind capture, disk formation, or a secondary stellar envelope. Furthermore, the model must extend beyond L2 and L3 to resolve any outflows that may form. Even with the computational savings of the TIME method, this method is only viable for the fastest of 3D hydrodynamical codes.

Computationally fast 3D hydrodynamics simulations are most often performed using smoothed particle hydrodynamics (SPH), which can capture the varied length scales of binary interactions efficiently. However, the computational savings of SPH is lost in simulations with complex boundary conditions and widely-varied density scales. As identified in \cite{dickson2024}, HMXB MT streams are fed simultaneously from fast diffuse gas on the stellar surface and slow dense gas from deeper within the envelope. Modeling the relative contributions of these flows self-consistently in SPH would ultimately require a greater computational cost than could be achieved with a grid-based approach of comparable effective resolution.

Instead, we have chosen the grid-based binary hydrodynamics code \vhone for our dynamical models \citep{vh1}. The modern implementation of \vhone, described in detail in \cite{blondintaylor}, expands the method of \cite{vh1} into 3D with two sets of partial spherical grids in the yin-yang configuration \citep{yinyang}. It solves the Euler equations for an ideal gas using a Lagrangian-Remap implementation of the piecewise-parabolic method \citep{ppm}. \vhone is able to resolve a finer effective resolution than its total number of zones suggests due to its double yin-yang grid \citep{yinyang}. This grid geometry combines two sets of two overset partial spherical grids to compose a high-resolution accretor-centered grid embedded within a low-resolution donor-centered grid \citep{blondintaylor}. 

While \vhone is among the most efficient 3D codes for global models of interacting binaries, the most effective modern version still requires $\unsim5,000~\text{cpu-hrs/orbit}$ with a manually-tailored grid domain \citep{time}. We have improved the efficiency of \vhone to $30-80~\text{cpu-hrs/orbit}$ through significant changes to the resolution, grid optimization, and computational domain.

Previous simulations with \vhone (e.g. \citealp{blondin2012hoyle,dickson2024,time,blondintaylor}) generally employ a $\unsim1^\circ$ angular resolution ($10^7$ zones), comparable to the highest effective resolution in use for 3D RLOF models. Under best-case circumstances, we have found global $\unsim1^\circ$ \vhone grids to require one orbit of relaxation at higher overflows and scale at $51\%$ efficiency to 1024 processors with respect to the baseline 128 processors used for high-resolution \vhone models. This equates to one evolutionary step per $20~\text{hours}$ of wall-time; at such a rate, extended overflow phases may require years to model. 

It is therefore imperative to limit the overall resolution of each 3D model in order to effectively execute hundreds or thousands of serial simulations. We have opted for a reduced resolution of $\unsim4^\circ$. The radial resolution has been likewise reduced, maintaining near-square zones in the wind regime and two pressure scale heights per zone within stellar envelopes. This lowered resolution, further tested in Section \ref{ssec:3dres}, enables a significant amount of acceleration to occur within a single Courant step \citep{courant1928cfl}; we reduce the Courant parameter to $0.25$ to compensate.

\subsubsection{Automatic Grid Optimization}

When appropriately scaled to the geometry of a particular system, the unique grid architecture of \vhone enables precise modeling while minimizing computational artifacts and computational cost. In past applications, it has always been a priori preferred that the grid architecture be manually tailored to each modeled system; in TIME series, an automatic approach is required. 

To adapt the 3D grid consistently across thousands of evolutionary timesteps without manual interface, we have modified \vhone to compute, iteratively optimize, and automatically debug the grid geometry. In addition to a specified grid resolution and inner boundary thresholds for each component, this computation is made with respect to the interior local scale heights of stellar components as well as the binary separation and mass ratio between components. The grid is iteratively tested and rescaled to ensure the donor grid contains the accretor grid and the outer Lagrange points and the accretor grid captures the majority of its Roche lobe without extending significantly into the donor. The local resolution and squareness of zones is weighed and adjusted at the donor surface and the interface between grids. This adaptive grid becomes particularly critical in very close binaries with either a large accretor or extended accretion disk, requiring the high-resolution accretor-centered grids to span a large portion of the overall grid. 

Unlike the method of \cite{dickson2024}, we scale the radial extent of zones below the stellar photosphere proportional to the local scale height to maintain a consistent resolution with respect to the local pressure scale height. For our $4^\circ$ grid, this equates to two scale heights per zone. This reduces excess computation as the deep envelope consistently forms steadier, slower flow structures than the photosphere, particularly near the inner boundary. To maintain a spherical coordinate system, this scaling cannot precisely match at each point within tidally-distorted stellar envelopes. Furthermore, the envelope structure will change across the course of the simulation and a fixed grid is preferable for computational cost. As such, we further modify our scaling method to minimize the gradient of zone height from one zone to the next. This generates a smooth quasi-exponential gradient in zone height.

\subsubsection{3D Computational Domain} \label{ssec:compdomain}

The computational domain of our simulation spans two gridsets, a low-resolution set centered on the donor and a high-resolution set centered on the accretor. 

The donor grid spans from approximately $0.8R_*$ to $1.8\times$ the binary separation, though with the adaptive grid scaling, this varies slightly from one 3D model to the next. In the case where one or more outer Lagrange points are greater than $1.8\times$ the binary separation, the automatic grid solution extends the outer boundary to include them. The accretor grid spans from $0.8R_*$ to L1 in the case of a stellar accretor; for compact object accretors, such a boundary would require additional physics to compute.

The area immediately surrounding an accreting compact object may be subject to extreme magnetic fields and radiation, as well as relativistic effects \citep{accretionpower}. While these effects can be critically relevant to observational signatures of HMXBs, precisely modeling compact object accretion requires computationally-intensive simulations even with much smaller computational domains \citep{narayan2012grmhd,liska2018grmhd}. For sake of computational efficiency, we define all material that falls within $1R_\odot$ of a compact object accretor to be accreted. This domain is expected to effectively resolve both wind-fed and disk-fed MT mechanisms, as it falls well below the Bondi-Hoyle-Lyttelton (BHL) radius and disk circularization radius of most massive binary systems. 

The BHL radius of a wind-fed accretor defines the maximal radius at which incident wind is gravitationally captured. This can be expressed as a function of the accretor mass, $M_a$, and relative wind velocity as

\begin{equation}
    R_{BHL}=\frac{2GM_a}{v_{local}^2}\geq \frac{2GM_a} {v_{\infty}^2} \quad .
\end{equation}

In HMXBs, $R_{BHL}\unsim 1-10 ~R_\odot$ at typical mass ratios, above our chosen boundary \citep{bondihoyle,demircan1995massratiodistribution}. This places our boundary safely within the capture region of systems undergoing direct wind accretion.

When accretion is instead mediated by a disk, the capture region is defined by the circularization radius \citep{el2019disks}. This is the radius at which material in an accretion disk circulates at a Keplerian velocity; accreting material below this point cannot maintain a stable orbit. This is typically expressed as 

\begin{equation}
    R_{circ}=\frac{l^2}{GM_a} \quad ,
\end{equation}

where $l$ is the specific AM of inflowing material. The circularization radius in HMXBs is typically $R_{circ}\unsim\text{a few}~R_\odot$ \citep{accretionpower,inoue2021accretion}.

\subsection{Informing 1D with 3D} \label{ssec:3Dto1D}

To inform the subsequent evolutionary step, each 3D model must be evolved to form a well-defined steady state. We achieve this with two orbits for relaxation and three orbits for steady-state extraction, with three additional orbits of relaxation given to the first model in each RLOF phase (see Section \ref{ssec:1Dto3D}).

While we could take a single orbit to characterize the steady state, lengthening each 3D model counterintuitively shortens the overall TIME series. The span of each evolutionary timestep depends in part on how significantly the system has changed from the previous step. The computational expense of the TIME method remains the 3D steps, but increasing this expense is still worthwhile to ensure smooth and precise 3D data. As shown in Table \ref{tab:orb}, spanning additional orbits increases precision significantly, leading to drastically shorter evolutionary timesteps. In preliminary testing, we found this increased precision could lengthen evolutionary timesteps by up to two orders of magnitude, reducing the total number of 3D steps required to span the studied MT phase.

\begin{figure*}
    \centering
    \includegraphics[width=1\linewidth]{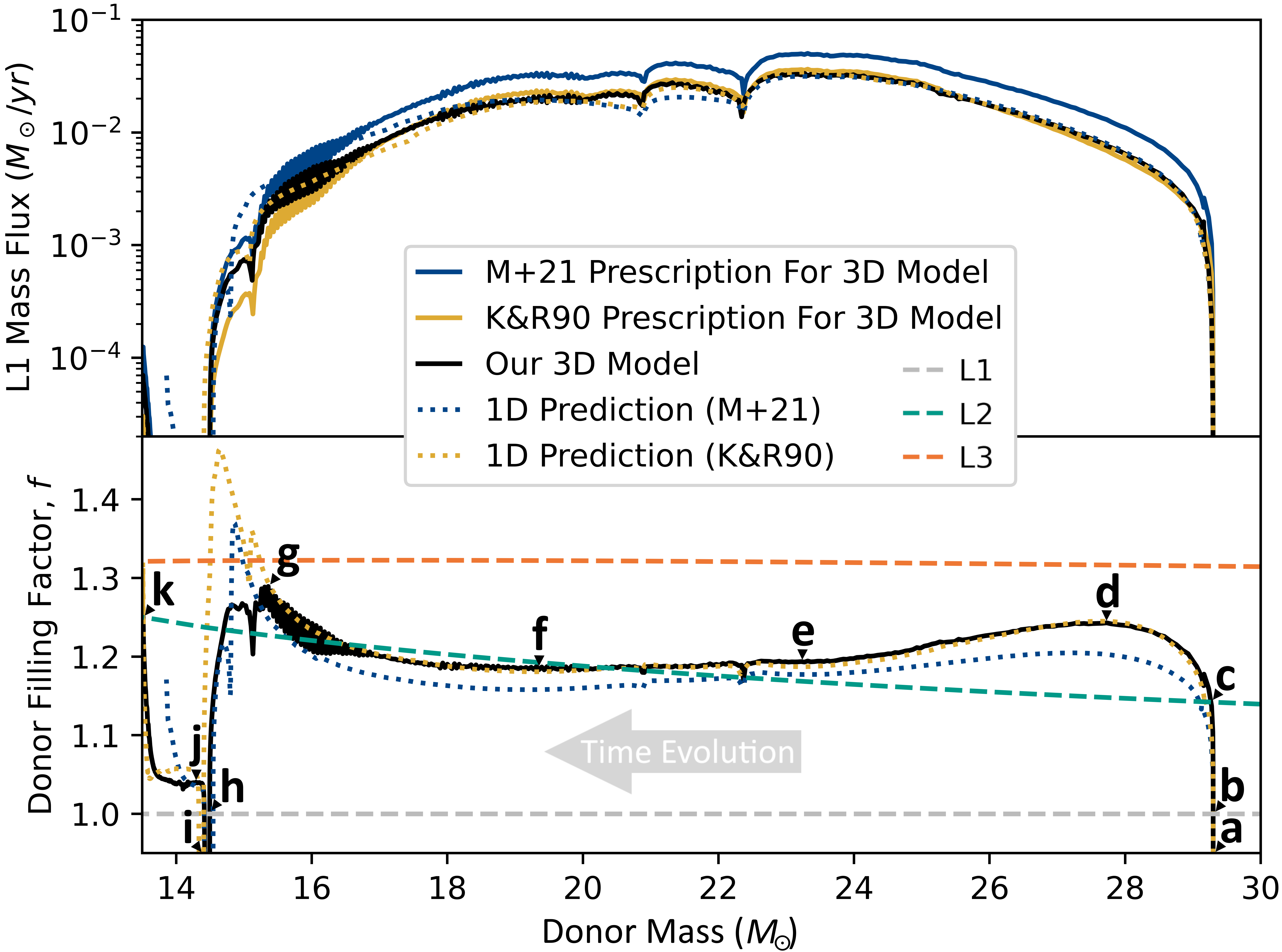}
    \caption{Evolution of Case B MT in our test BBH progenitor. The colored solid lines represent the analytical prescriptions of \cite{kolb&ritter} and \cite{marchant2021roleofMT} for each evolutionary step of our 3D model. Dotted lines indicate the evolution obtained with 1D \mesa using those prescriptions in lieu of hydrodynamics. Dashed lines approximate the effective radii of L1-3, computed with the methods of \cite{eggleton1983}, \cite{marchant2016l2}, and \cite{marchant2021roleofMT}, respectively. Letters correspond to the evolutionary stages visualized in Figure \ref{fig:evo3d}. The envelope stripping phase (near point g) is resolved in detail in Figure \ref{fig:tconv}.}
    \label{fig:evovkolb}
\end{figure*}

Across this steady state sample, we may extract time-averaged fluxes across key system contours. These include the mass flux through L1, L2, and L3 of all material above wind densities, their associated AM, and the mass and AM loss of both the donor wind and the system as a whole\footnote{These values are taken at the outer bound of the simulation, which is scaled to just beyond the outer Lagrange points. Some material taken to be outflowing in this model might instead be retained by the system if a higher outer bound was taken.}. For each Lagrange point, we extract net fluxes from a $45^\circ$ spherical cap passing through the Lagrange point and centered about the donor while excluding material below the photospheric density. Any or all of these parameters can then be used to inform the subsequent evolutionary step; to limit the scope of our first test of the framework, we feed back only the L1 mass flux in this work. We track, but do not feed back, the fluxes of the L2/L3 outflows to estimate their impact if taken into account in future simulations.

This flux is set as an explicit fixed parameter in the following \mesa step. In 1D models, \mesa approximates this flux with an analytical MT prescription across implicit time evolution \citep{mesa3binary}. Other than this change, all other assumptions of the \mesa model are retained and evolved from the prior 1D profile. 

\section{Stable Case B MT in a Merging BBH Progenitor} \label{sec:test}

\subsection{Test System} \label{ssec:testsys}

We apply our 3+1D framework to BBH progenitor near the boundary of stable mass transfer as predicted by \cite{marchant2021roleofMT}. We model this system during the HMXB phase, the final stage of stable evolution prior to BBH formation. Our test system is initially composed of a $9.3~ M_\odot$ black hole accretor and a $30~ M_\odot$ donor star initiated as a zero age main sequence star at metallicity $0.1~Z_\odot$. They begin in a circular orbit with a period of 10 days. This system is evolved both by a standalone 1D \mesa evolution following the mass loss prescriptions of \cite{kolb&ritter} and \cite{marchant2021roleofMT} and by our 3+1D hybrid code across two complete phases of Case B MT, early and late. Early Case B MT, triggered by the thermal-timescale ($\unsim \text{kyrs}$) expansion that occurs after the end of the main sequence, is generally characterized by highly-extended donors that may overflow their L2 and L3 equipotentials. Our model subsequently undergoes a late Case B MT phase triggered by gradual expansion during Helium burning on nuclear timescales ($\unsim 100~\text{kyrs}$) spanning much of the core Helium burning phase. We end our simulation at core Carbon depletion; we therefore conclude this second Case B phase but neglect consideration of a subsequent Case C phase, instead taking the donor to undergo direct collapse into a black hole of the same mass it had at the core Carbon depletion to estimate BBH merger time.

Though late Case B donors frequently remain overflowing at and beyond core Helium depletion, this threshold is considered the end of Case B MT and the beginning of a subsequent Case C phase which may differ significantly in timescale.

Our 3+1D TIME series follows the Eddington-limited MT assumption to test the validity of that approximation. This approximation defines all flow between components to occur through L1, with no loss from the system through L2/L3 outflow streams. We therefore only feedback the L1 mass flux from 3D to 1D, as described above, though we still track other system fluxes in great detail. Most notably, the mass and AM loss through the L2 and L3 points. These fluxes are neglected or parameterized by many binary evolution and population synthesis models (e.g. \citealp{king1999eddington,schneider2024presupernova,picco2024forming}). Though there is a consensus these flows play an important role, they have remained largely unconstrained beyond dynamical-timescale simulations \citep{lu2023rapidmt,scherbak2026radiativel2}. By employing 3D hydrodynamics throughout the transfer phase, we are able to test the overall significance of non-conservative outflow streams and the specific AM they carry during stable MT.

Systems similar to the HMXB we describe have generally undergone one or more prior MT phases in which the now-donor accreted material from its companion star before its collapse into a black hole, altering its composition \citep{xu2026stablemt}. A complete picture of the impact of 3D MT from zero age main sequence to BBH formation remains beyond the scope of this work. Future work may benefit from exploring the cumulative effect of these phases modeled in 3D.

\subsection{Results} \label{ssec:results}

\begin{figure}
    \centering
    \includegraphics[width=1\linewidth]{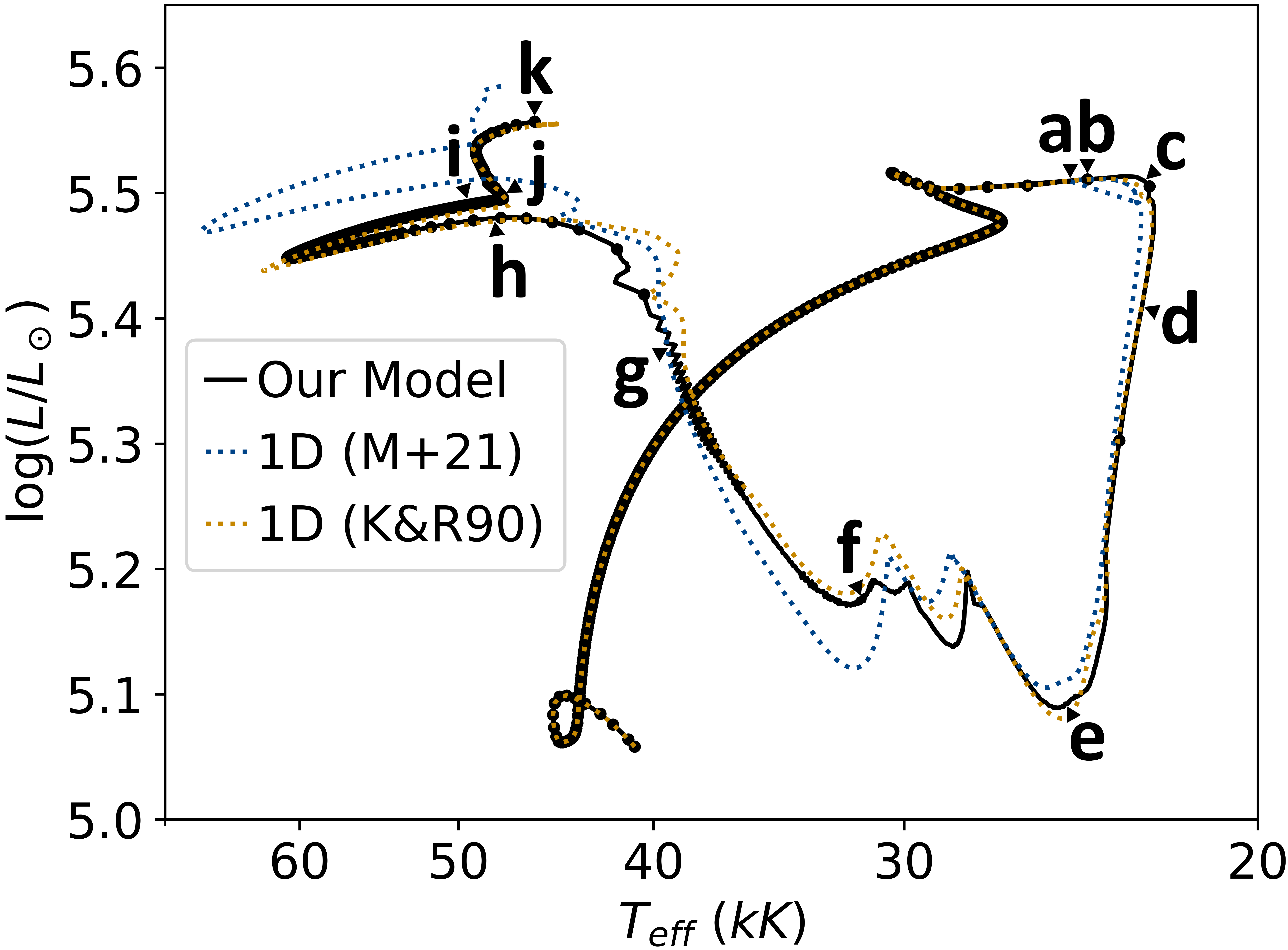}
    \caption{A Hertzprung-Russell Diagram charting the evolution of our test BBH progenitor. Dotted lines indicate the evolution obtained with 1D \mesa using the prescriptions of \cite{marchant2021roleofMT} and \cite{kolb&ritter} in lieu of hydrodynamics. Letters correspond to the evolutionary stages visualized in Figure \ref{fig:evo3d}. Black dots are spaced at 500 yr increments to visualize the rate of evolution.}
    \label{fig:hrd}
\end{figure}

\begin{figure*}
    \centering
    \includegraphics[width=1\linewidth]{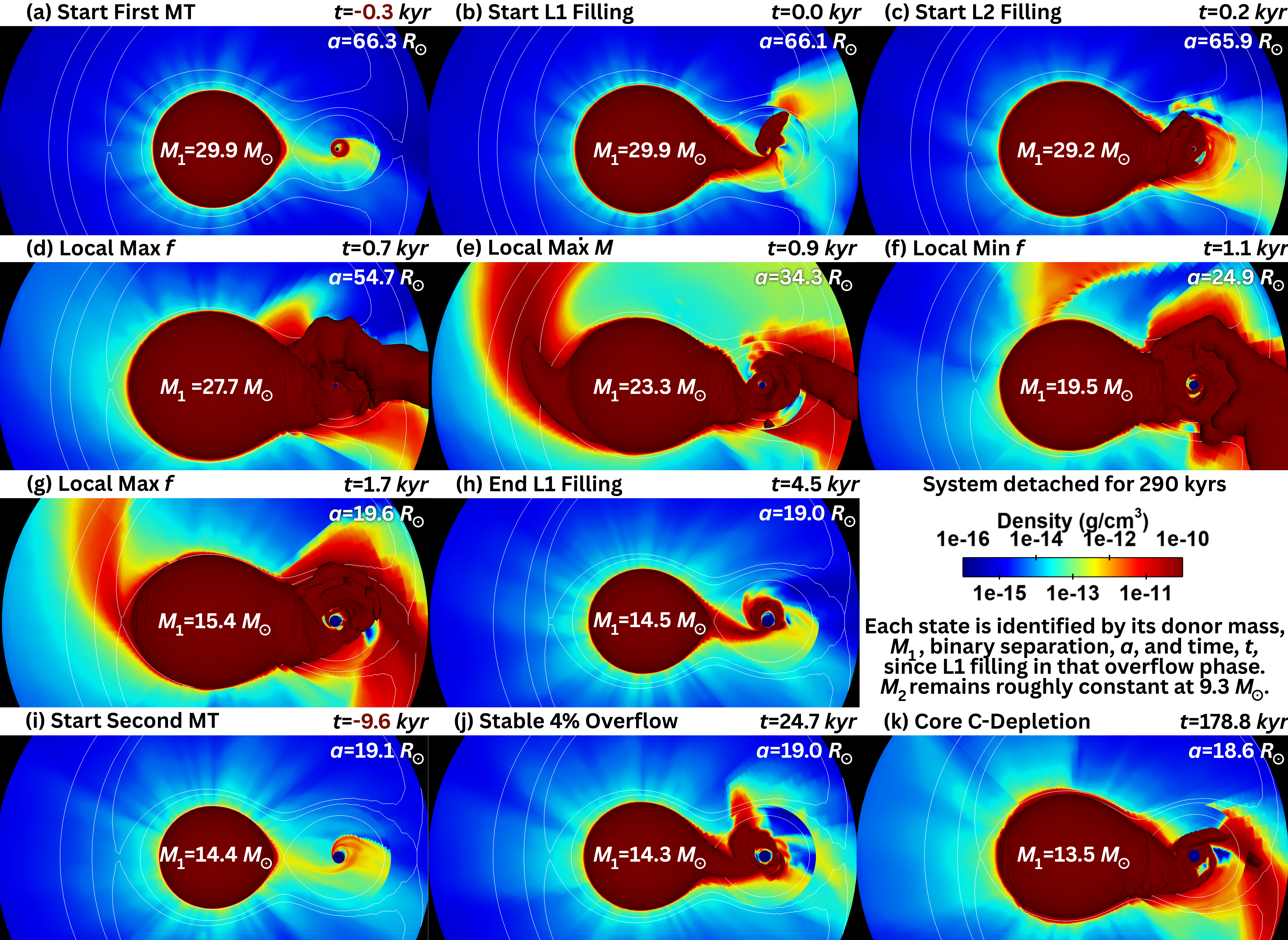}
    \caption{3D Dynamics of Case B MT across evolutionary time in our test BBH progenitor. The ecliptic plane is colored by density. A density isosurface is applied to visualize the off-ecliptic system geometry. Subplots are scaled by their binary separation for visibility. See also \href{https://zenodo.org/records/24324841}{Movie 1}.}
    \label{fig:evo3d}
\end{figure*}

Our 3+1D test case diverges from the 1D model after the end of the main sequence, when the donor fills its Roche lobe by $f=R_*/R_\text{RL}=0.95$ and triggers the first 3D model in the series. From that point, the system overflowed for $5$ kyr, stripping the donor envelope across a thermal-timescale early Case B MT phase. During this phase, the last of the donor envelope rapidly expands and becomes increasingly diffuse as progressively deeper layers are stripped away until the core remains as a detached stripped star. After detachment, the donor remained well-contained by its Roche lobe for $290$ kyr before expanding in the latter half of core Helium-burning. The star then underwent a nuclear-timescale (late Case B) MT phase that spanned an additional $189$ kyr before core Carbon depletion. This evolution spanned $1089$ evolutionary steps, including $574$ 3D models of five orbits each and $515$ steps of detached evolution. Two of our 3D models, the first in each MT phase, were evolved for a orbits as noted in Section \ref{ssec:relax}. In total, our model required $1.8\times10^5~\text{cpu-hrs}$ at an average of $62~\text{cpu-hrs/orbit}$.

We illustrate these two MT phases in Figures \ref{fig:evovkolb}, \ref{fig:hrd} $\&$ \ref{fig:evo3d}. We additionally provide an animated view of the first MT phase in \href{https://zenodo.org/records/24324841}{Movie 1}. This movie is scaled by evolutionary timestep rather than time to better visualize rapid transitions. Figure \ref{fig:hrd} additionally visualizes the prior 1D evolution across the main sequence, though physical systems of this kind would be expected to have undergone a prior phase of MT driven by the progenitor of the black hole not modeled in this work.

The first 3D model has an approximate BHL radius of $4.3~R_\odot$ and circularization radius of $4.6~R_\odot$. As the orbit hardens across the MT phase, the BHL radius increases while the circularization radius remains roughly constant. Both remain well above the compact object inner boundary of $1~R_\odot$, providing theoretical support to the validation performed in Section \ref{ssec:compdomain}. 

The MT process holds to be self-regulating, maintaining much the same transfer rate regardless of the overflow mechanism required to achieve it. The L1 MT rate obtained through our 3D hydrodynamics simulations was closely approximated by the prescription of \cite{kolb&ritter} for the same overflow conditions until the final stages of envelope stripping. The more updated prescription of \cite{marchant2021roleofMT} consistently overestimated L1 mass loss, but remained accurate to within a factor two across the majority of the evolution. Our data suggests existing prescriptions (particularly that of \cite{kolb&ritter}) closely predict real L1 mass loss rates in massive Case B systems, with an important exception. In the final stage of thermal MT, between points g and h in Figure \ref{fig:evovkolb}, both prescriptions diverged significantly, predicting more substantial outflows than our model suggests.

Our 3D models reveal substantial but not dominant outflows through the L2 and L3 regions. The first MT phase rises initially to exceed the L2 potential (Figure \ref{fig:evo3d}c) and form a strong L2 stream. Our 3+1D TIME series does not account for losses through this stream, but AM loss through Eddington-limited accretion rapidly shrinks the system from this point, quickly lowering the binary separation to a third of its initial value. Simultaneous to this change in orbital geometry, the donor envelope begins to be stripped away at rates quickly exceeding $10^{-2}~M_\odot/\text{yr}$. Between these two effects, we see a dense stream form at L3 (Figure \ref{fig:evo3d}e) despite L3 underflow. This stream is slowly cut off (Figure \ref{fig:evo3d}f) until re-emerging in the later stages of envelope-stripping (Figure \ref{fig:evo3d}g). 

1D models relying on any of the mass loss prescriptions of \cite{kolb&ritter}, \cite{marchant2021roleofMT}, or \cite{cehula2023mtprescription} predict significant L3 overflow in rapid MT similar to that undergone in our model (point g in Figure \ref{fig:evovkolb}). Our 3D model differed by remaining well-contained by L3, growing in radius during envelope stripping to a peak of $\unsim1.29~R_\text{RL}$, well below the $\unsim1.37~R_\text{RL}$ predicted by \cite{marchant2021roleofMT} predicted and $\unsim1.46~R_\text{RL}$ predicted by \cite{kolb&ritter}. \cite{cehula2023mtprescription} generally predict higher overflows than \cite{kolb&ritter} and further suggest these overflows could trigger unstable MT into a common envelope; the relatively low overflow of our 3+1D simulation avoided that possibility.

Nearing the end of the early Case B phase, MT slows and the donor envelope is stripped away until detachment (Figure \ref{fig:evo3d}i). After remaining detached for $290$ kyr, the donor re-expanded into a late Case B phase of nuclear-timescale MT at low ($\unsim4\%$) overflow (Figure \ref{fig:evo3d}j). During this phase, the donor never formed an L3 stream, but dense L2 flows were achieved far below the L2 equipotential. These flows may have been aided by the short period and high relative contribution of the wind, which has been previously suggested to enhance L2 mass loss in similar HMXBs \citep{dickson2024}. This stream was significantly outweighed by the losses of Eddington-limited accretion, but may be dense enough to potentially occlude X-rays and impact observations across nuclear timescales.

Following the Eddington-limited transfer assumption, our model differed by only $0.002~M_\odot$ from the total mass loss prediction of the 1D model following the prescription of \cite{kolb&ritter}. We did not account for outflows in this mass loss; however, our 3+1D model suggests they would not have significantly impacted the final donor mass (see Table \ref{tab:out}). If we assume the donor star to undergo direct collapse without mass loss immediately upon core Carbon depletion, our model predicts the formation of a BBH of masses $13.50+9.39~M_\odot$ and separation $18.62~R_\odot$, which would merge due to gravitational wave emission after $6.223~\text{Gyr}$. This amounts to a $<1\%$ difference from the merger time one would obtain for the 1D model following \cite{kolb&ritter} under the same assumption ($6.200~\text{Gyr}$).

\begin{table}
    \caption{Relative losses through outflow streams as compared to the total losses from the system through Eddington-limited re-emission.}
    \centering
    \begin{tabular}{cccc}
    \hline\hline
        Outflow & MT Phase & Mass Loss & AM Loss \\
        & & ($\%$ of $\dot{M}_{\text{1D},tot}$) & ($\%$ of $\dot{J}_{\text{1D},tot}$) \\
        \hline
        L2 & 1 & $1.2\%$ & $3.8\%$ \\
        L2 & 2 & $0.3\%$ & $1.0\%$ \\
        L3 & 1 & $0.004\%$ &$0.01\%$\\
        L3 & 2 & - & - \\
        \hline
    \end{tabular}
    \label{tab:out}
\end{table}

\begin{figure}
    \centering
    \includegraphics[width=1\linewidth]{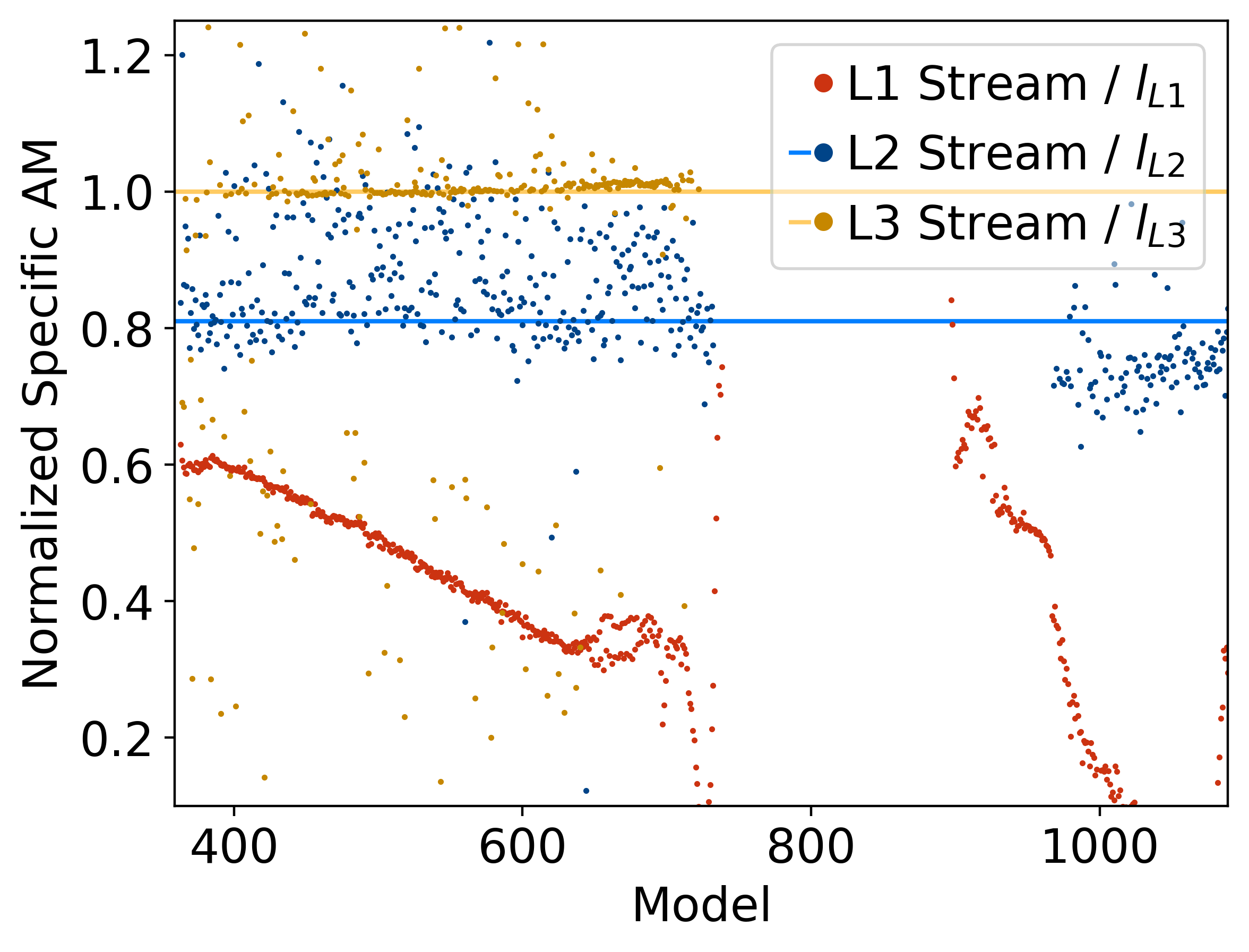}
    \caption{Specific angular momenta of tidal streams normalized to circular orbits through their respective Lagrange points. Constant fit lines are additionally provided for L2 and L3. Model 359 corresponds to the start of the first MT phase.}
    \label{fig:specangmom}
\end{figure}

We also tracked the specific AM of the three streams across both MT phases. We plot these normalized against the AM of circular orbits at their respective Lagrange points in Figure \ref{fig:specangmom}. We find the specific AM of the L2 and L3 streams to be quite consistent. The L3 stream closely follows that of its Lagrange point in most models, but as the stream formed and detached, it experienced substantial outliers. It is unclear whether these outliers are physical or the result of insufficient model relaxation. The L2 stream had a moderate dispersion, but remained fairly constant somewhat below the AM of its Lagrange point. We obtain a mass-weighted average specific AM of $0.810\pm0.007 ~ l_{L2}$ in the L2 outflow stream across both transfer phases.

The L1 stream did not maintain constant specific AM, instead changing substantially between simulations. In both MT phases, the L1 stream began initially above half the specific AM of its Lagrange point and steadily declined. After core He depletion, we see a rapid rise in the final models of our series (Figure \ref{fig:specangmom}). This may suggest that the specific AM of the L1 stream depends substantially on the envelope structure\footnote{Though this stage also comes with a rapid rise in overflow that may be partially responsible for the rise in specific AM, we note that the first MT phase saw specific AM decline across increases in overflow.}. This may significantly impact the accretion flow, e.g. in systems where the change of circularization radius might transition a star+star binary between disk accretion and direct impact.

Seen in both Figure \ref{fig:evo3d} and \href{https://zenodo.org/records/24324841}{Movie 1}, the accretion disk was unable to remain compact and thin at high overflow. Though the disk was able to retain a thin structure at low overflow, our model may artificially inflate the disk due to relatively low angular resolution. Further work is required to assess the validity of the thin disk model in similar Case B systems.

\section{Discussion} \label{sec:disc}

In addition to its relevance to a range of ongoing research (e.g. \citealp{marchant2021roleofMT,vanson2022stablemt,lu2023rapidmt,xu2026stablemt}), we chose our test case to stress test our 3+1D TIME framework on multiple benchmarks. While observed HMXBs tend to be at very low overflow, the chosen early Case B phase was predicted by 1D models to overflow both L2 and L3, generating highly 3D outflow structures and a large accretion disk as it strips the donor envelope. The extreme overflow of envelope stripping provides a strong test of our 3D mapping from the 1D profile (evaluated in Section \ref{ssec:1dres}) while the highly sensitive response of stripped envelopes tested the limits of our 1D time resolution (Section \ref{ssec:1dt}). The complex structure and high mass of the disk and outflows present worst-case conditions for model relaxation, testing our chosen relaxation time and disk-initialization method and 3D resolution (Sections \ref{ssec:relax} $\&$ \ref{ssec:3dres}). Highly-overflowing HMXBs also present the greatest range of length and timescales of any MT systems, pushing the computational domain. This provided the ideal ground to test our boundaries about the compact object, within the deep donor envelope, and out beyond the outer Lagrange points (Section \ref{ssec:3dbound}). 

\subsection{1D Spatial Resolution} \label{ssec:1dres}

\begin{figure}
    \centering
    \includegraphics[width=1\linewidth]{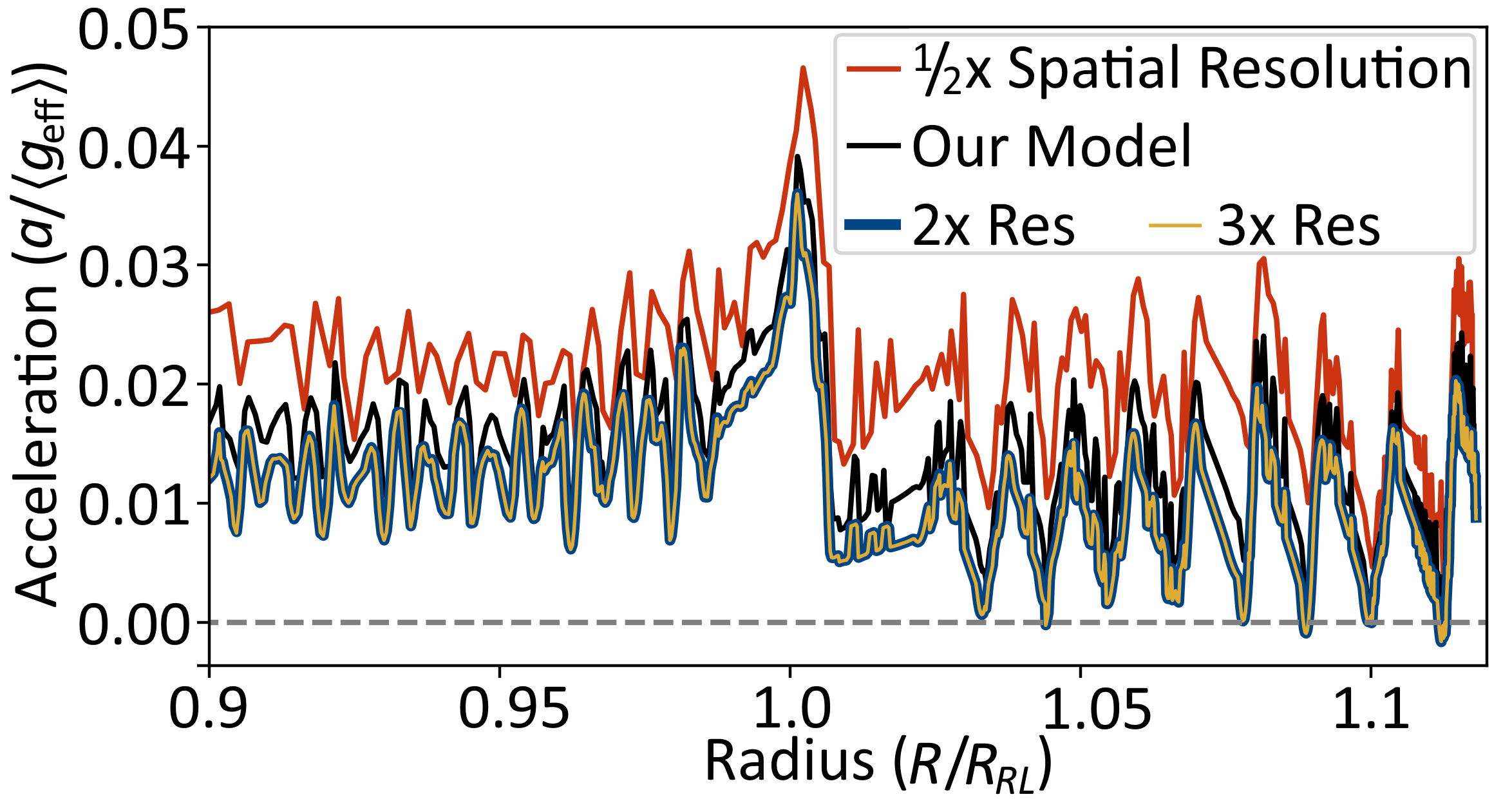}
    \caption{Deviation from hydrostatic condition with respect to radius across a range of 1D mesh resolutions. All four models utilize the same pre-computed geometry table.}
    \label{fig:1dmesh}
\end{figure}

In Figure \ref{fig:1dmesh}, we test the impact of our 1D mesh resolution by evaluating the acceleration given by Equation \ref{eqn:accel}; this can be computed directly without mapping into 3D hydrodynamics. We take this comparison at a moderate $12\%$ overflow (between b and c in Figure \ref{fig:evovkolb}). This choice places the midline of our envelope profile (which descends to $0.8~R_*$) approximately at the Roche lobe radius, such that we can test the hydrostatic condition in both the bound and unbound regimes. 

Figure \ref{fig:1dmesh} shows our employed resolution to be a noticeable improvement over the standard resolution. Future models may find some benefit by an additional doubling of the mesh resolution\footnote{To $\texttt{mesh}\_\texttt{delta}\_\texttt{coeff} = 0.1$.}; while this further refinement offers only small advantages, the costs of refinement (both in data storage and increased \mesa runtime) are negligible compared to the costs of the accompanying 3D simulations. Future models may benefit from computing a finer grid for interpolating the 1D geometry as described in Section \ref{ssec:mesa}. However, the accelerations induced by our method are on the order of a few $\%~g_{\rm eff}$ and should not significantly impact results obtained.

\subsection{1D Time Resolution} \label{ssec:1dt}

\begin{figure}
    \centering
    \includegraphics[width=1\linewidth]{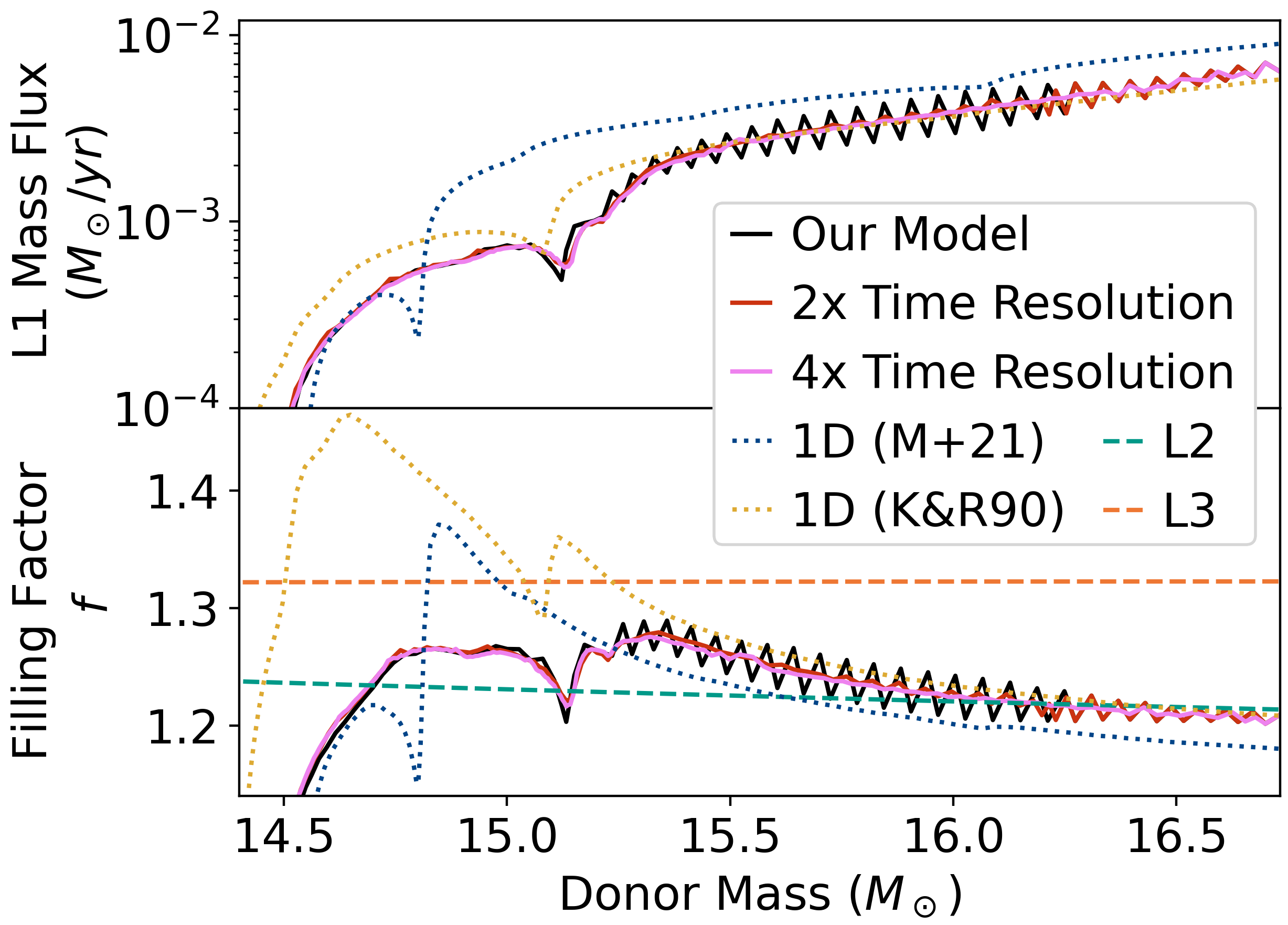}
    \caption{Evolution of filling factor and MT rate at a range of time resolutions. Dotted lines indicate the evolutionary path \mesa would take without feedback from hydrodynamics at the original time resolution using the prescriptions of \cite{kolb&ritter} and \cite{marchant2021roleofMT}, respectively.}
    \label{fig:tconv}
\end{figure}

We also validated our TIME series against variations in the effective \mesa time resolution (Figure \ref{fig:tconv}). To test the regime most vulnerable to variations in evolutionary timestep, we performed this comparison on the final stage of envelope-stripping at the end of the early Case B phase. We evolve the same initial system state across TIME series at $2\times$ and $4\times$ the effective time resolution of our model. In Figure \ref{fig:tconv}, we track each simulation by its filling factor and MT rate through L1. 

Figure \ref{fig:tconv} demonstrates an oscillatory instability which arose in our model. As the last of the envelope is stripped away, the radius of the donor becomes increasingly sensitive to MT, and our model time resolution was insufficient to smoothly evolve that transition. The artificial oscillations we observe during envelope stripping amount to as much as a $3.4\%$ delta in overflow (from $21.1\%$ to $24.5\%$), inducing as much as a $50\%$ variation in the L1 MT rate between successive steps. Similar but much smaller oscillations occur elsewhere in our dataset, such as seen in Figure \ref{fig:evovkolb} at a donor mass of $18~M_\odot$. Despite these fluctuations, our model agrees closely with the evolutionary paths determined by the two finer time resolutions (Figure \ref{fig:tconv}). All three models reach $14.5~M_\odot$ at an overflow of $7\pm1\%$ with $10\pm1\%$ of their envelope remaining. 

The models with $2\times$ and $4\times$ effective time resolution required $23\%$ and $96\%$ more 3D steps to achieve the same endstate along a smoother progression. This suggests future models studying sensitive systems may prefer to employ twice our time resolution\footnote{To $\texttt{time}\_\texttt{delta}\_\texttt{coeff}=0.5$} to improve precision without dramatically higher computational cost. The artificial oscillations became most significant with a very reactive envelope incurring high changes in overflow; future models may also benefit from a more targeted change to timestep controls, i.e. reducing the maximum change in overflow allowed between successive \mesa steps.

\subsection{Accretion Disks and 3D Relaxation Time} \label{ssec:relax}

We compare models both with and without disk initialization across twenty orbits to evaluate our model relaxation. We make this comparison at the first peak of overflow in the early Case B phase (Figure \ref{fig:evo3d}d). This represents the worst-case scenario for both approaches, because the rapid expansion of the donor just after the end of the main sequence generates a very large disk that changes significantly from one model to the next.

\begin{figure}
    \centering
    \includegraphics[width=1\linewidth]{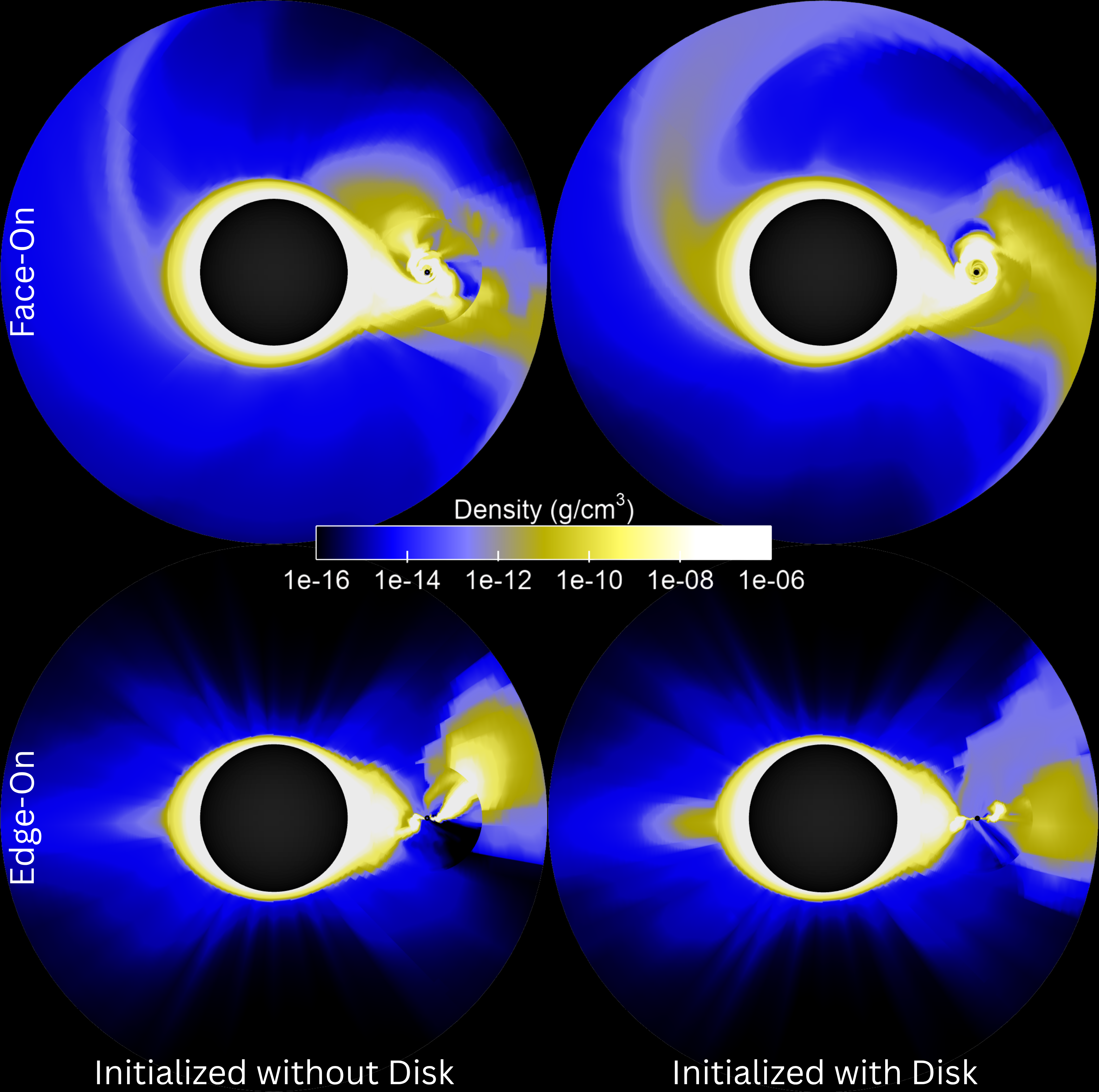}
    \caption{Persistent impact of disk-initialization after twenty orbits of relaxation.}
    \label{fig:diskval}
\end{figure}

Figure \ref{fig:diskval} demonstrates that even after twenty orbits, significant discrepancies remain between disk-initialized models and those initialized without a disk. The disk of the non-disk-initialized model has not yet resolved the high eccentricity and inclination that accompanied its formation. This results in more variable outflows as well as backflows onto the donor as disk material exceeds the accretor Roche lobe\footnote{These variable disk-formation flows may be physical only in MT phases that progress on a dynamical timescale.}. Disk-initialized models effectively remove these backflows, as shown in Figure \ref{fig:diskval}.

Despite these discrepancies in outflows and accretion disk structure, we found the L1 mass fluxes of these models converge towards a common steady-state, differing by only $0.5~\%$ in their respective steady-state mean L1 fluxes after sufficient relaxation time. 

As summarized in Table \ref{tab:orb}, initial model relaxation in disk-initialized models spans no more than $1-2$ orbits with subsequent changes generally shifting by $<5\%$. Based on this relaxation timescale, we take the first two orbits as sufficient relaxation. Once the model is relaxed, variance remains quite low, but occasional transient features vary the MT rate by tens of percent for a small fraction of an orbit. Some such features are seen in Figures \ref{fig:resol}, \ref{fig:donbound} $\&$ \ref{fig:accbound}. The effect of these transients is effectively mitigated by spans of three post-relaxation orbits. Table \ref{tab:orb} suggests our mean steady state values for the L1 mass flux consistently remain within a few percent of the true steady state values while being robust to transient behaviors.

\begin{table}
    \caption{Steady state of disk-initialized model at $24\%$ overflow.}
    \centering
    \begin{tabular}{cccc}
    \hline\hline
        Span & L1 Mass Flux & Error & Variance \\
        \hline
        ($Orbits$) & ($\log M_\odot/\text{yr}$) & ($\sigma_{\text{Long-Term}}$) & ($\sigma^2/\dot{M}_{avg}$)\\
        $0-1$ & $-2.128$ & $0.50$ & $16\%$\\
        $1-2$ & $-2.110$ & $1.35$ & $1\%$ \\
        $2-3$ & $-2.123$ & $0.03$ & $6\%$ \\
        $17-20$&$-2.123$ & $0.06$ & $1\%$ \\
        $2-20$ &$-2.123$ & $-$ & $2\%$ \\
        \hline
    \end{tabular}
    \label{tab:orb}
\end{table}

\subsection{3D Resolution} \label{ssec:3dres}

\begin{figure}
    \centering
    \includegraphics[width=1\linewidth]{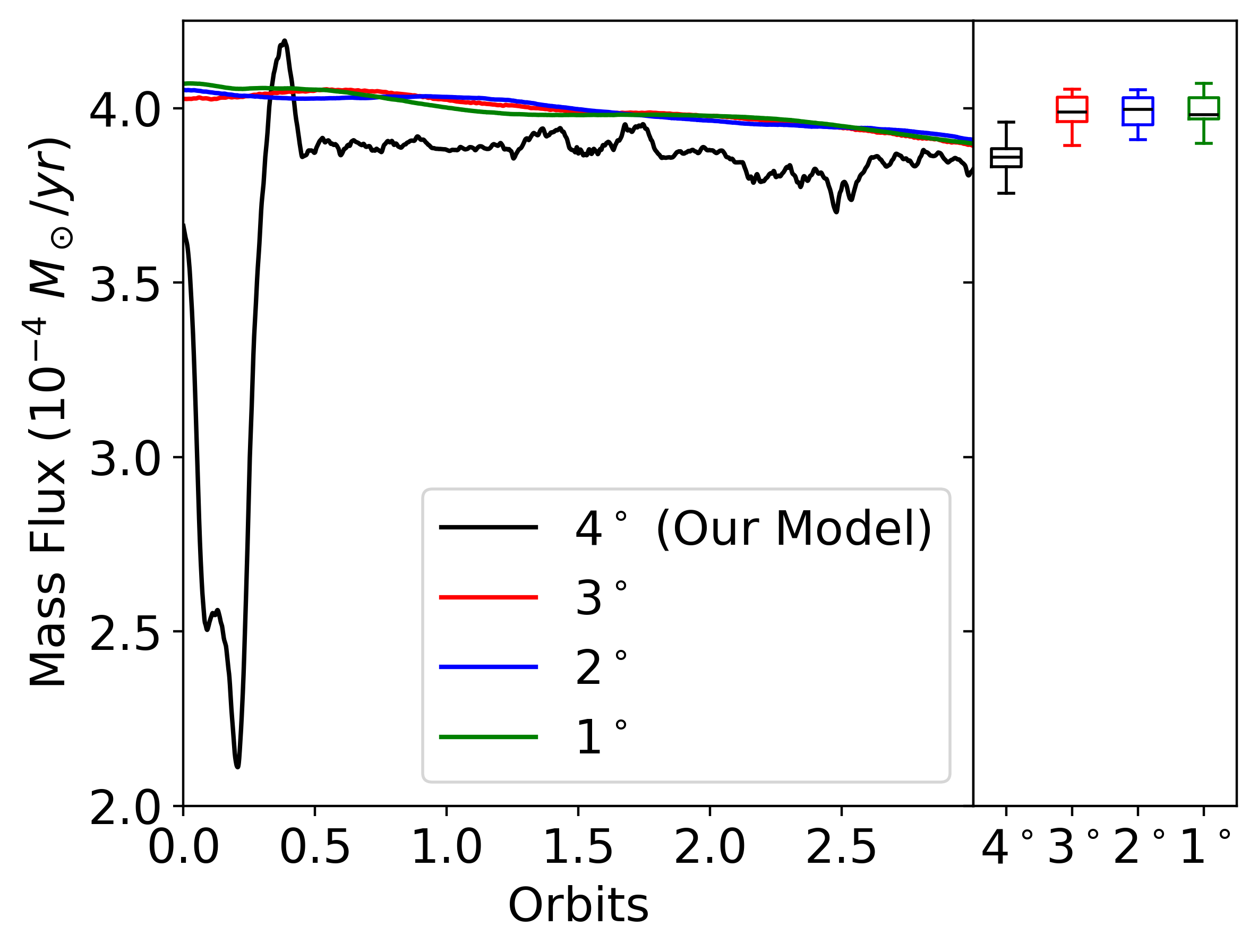}
    \caption{Steady state L1 mass flux at a range of grid resolutions. Each resolution is denoted by the size of zones in the angular dimensions. We maintain a Courant parameter of $0.25$ across all four simulations for consistency.}
    \label{fig:resol}
\end{figure}

We test our model at four grid resolutions (Figure \ref{fig:resol}). We perform this comparison at the beginning of early Case B MT with a filling factor $f=R_*/R_\text{RL}=1.12$, as done in Section \ref{ssec:1dres}. The dense envelope, rapid MT, and compact disk geometry of this stage present a strong test of low-resolution methods. 

While the $4^\circ$ model is significantly more variable than finer grids, both its variability and error with respect to the $1^\circ$ model remain quite small. The substantial agreement across significant variations in grid resolution shown in Figure \ref{fig:resol} leads us to conclude that our findings remained accurate to within $6\%$ of the highest resolution contemporary models under worst-case conditions. As we cannot conduct full TIME series at each resolution, we performed this comparison with non-disk-initialized models. As disk-initialized models suppress variability induced by disk relaxation (Figure \ref{fig:diskval}), this can be taken as an overestimate of the error between our model and equivalent high-resolution models.

Figure \ref{fig:resol} suggests a $3^\circ$ grid may provide a worthwhile improvement in precision for future studies, markedly reducing variability and substantially improving accuracy with respect to the $1^\circ$ solution.

\subsection{3D Boundaries} \label{ssec:3dbound}

\begin{figure}
    \centering
    \includegraphics[width=1\linewidth]{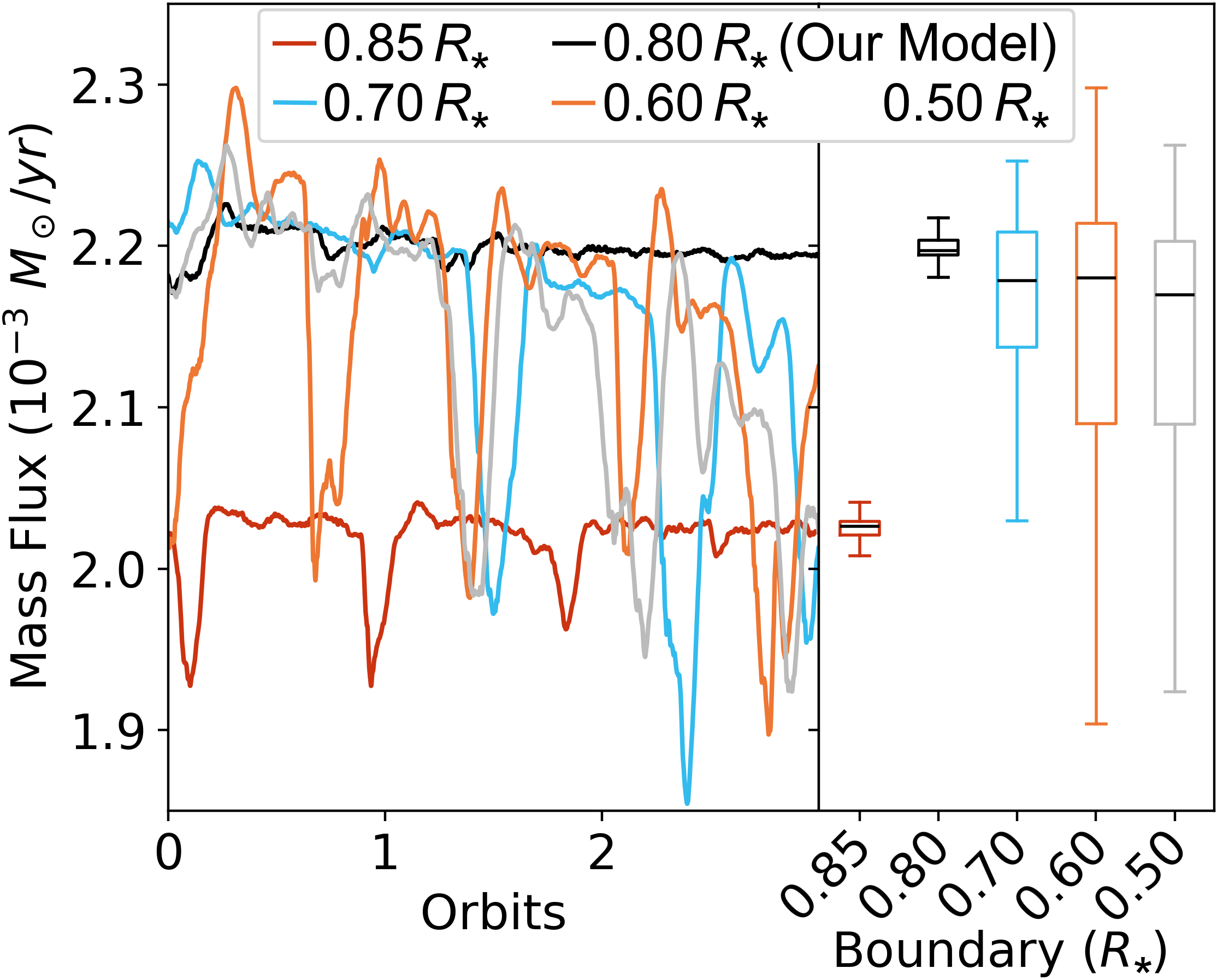}
    \caption{Steady state L1 Mass Flux at a range of donor inner boundary depths. Radii are given in units of the donor radius. Lower values indicate the simulation spans deeper into the donor envelope.}
    \label{fig:donbound}
\end{figure}

To ensure our model is robust to our choice of boundary conditions, we have performed two comparisons in opposite extremes of RLOF. 

\cite{dickson2024} suggests the donor inner boundary is most impactful upon the model in high overflow, when deep upwellings contribute to the MT stream. To test our model's sensitivity to the donor inner boundary, we perform a comparison at the peak expansion point during envelope-stripping (Figure \ref{fig:evo3d}g) across a range of boundary conditions. During this brief point, the envelope is drastically over-extended and diffuse, leading to extreme overflows, $29\%$ for this system. The diffuse surface of the envelope during this phase makes the effect of deep upwelling flows, and therefore the donor inner boundary, more significant than in any other stage of MT. 

Shown in Figure \ref{fig:donbound}, we find close agreement in mean steady state values between tested boundary depths. Lowering the boundary below $0.8~R_*$ introduced fluctuations to the mass flow through L1 resulting from turbulent flow structures in the deep envelope radially below the L1 point. However, despite the fluctuations induced, the mean mass flux through L1 remains within $3\%$ of the value obtained at our chosen boundary for even very deep boundary conditions, validating our choice of boundary (Figure \ref{fig:donbound}).

This extreme overflow provides a strong test for donor inner boundaries but is a rather weak test of compact object boundaries. High overflow generates a proportionally large-scale disk structure, moving the relevant dynamics further from the accretor. In contrast, barely-overflowing systems only form weak tidal streams and small-scale structures about the compact object. The streams of these $f\unsim1$ donors are most likely to be impacted by the choice of accretor inner boundary about a compact object. 

\begin{figure}
    \centering
    \includegraphics[width=1\linewidth]{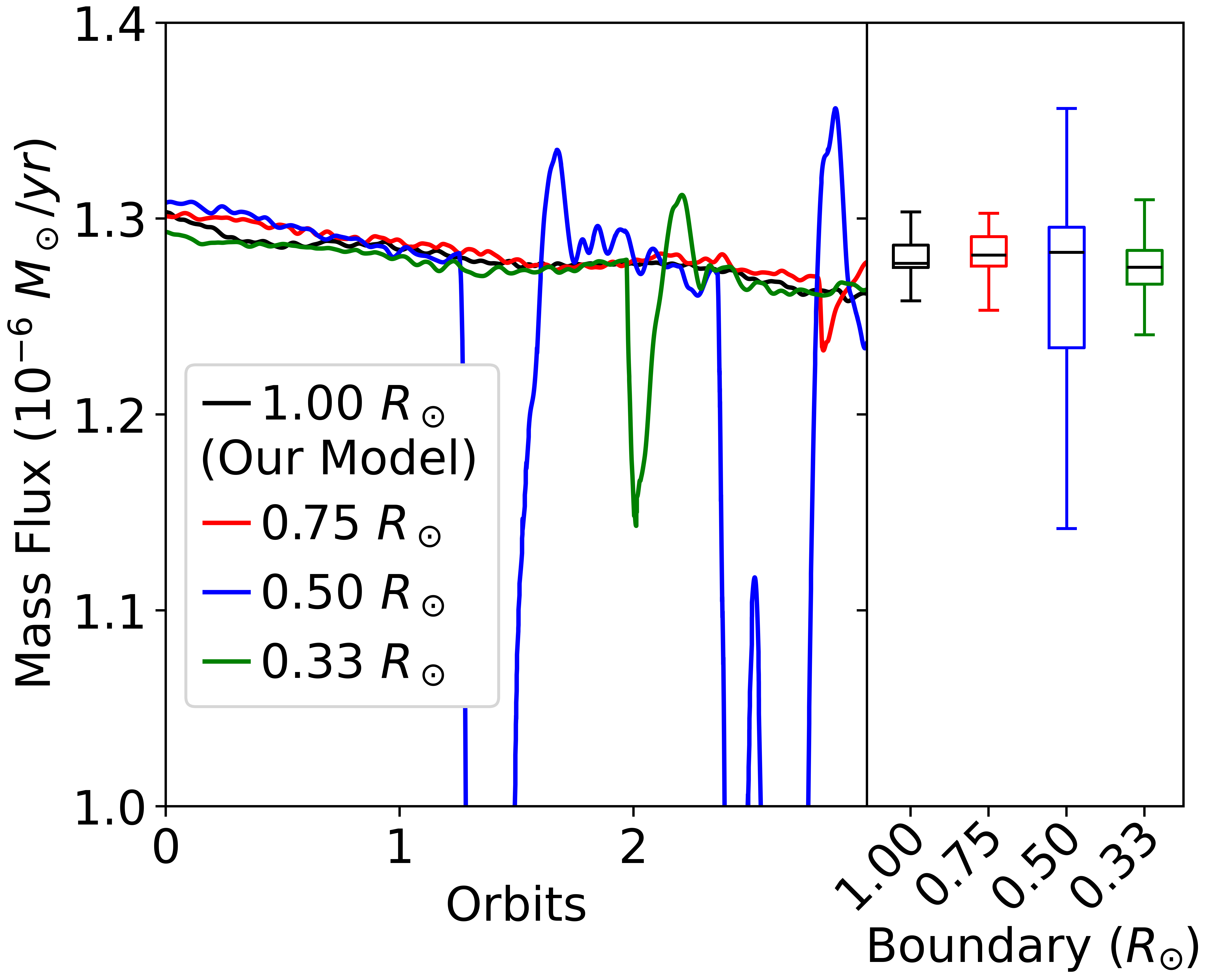}
    \caption{Steady state L1 Mass Flux at a range of compact object accretor inner boundaries. Lower values indicate the simulation extends closer to the compact object surface.}
    \label{fig:accbound}
\end{figure}

We tested the system during initial Roche filling (Figure \ref{fig:evo3d}b) a range of boundary conditions about the compact object accretor (Figure \ref{fig:accbound}). We find substantial agreement between our model ($R_{bound}=1.00~R_\odot$) and a boundary set up to three times closer ($R_{bound}=0.33~R_\odot$), which differed by an order in computational cost. We find that lowering the boundary closer to the compact object does impact the formation of wind-driven accretion structures, such as those that induced higher variability in the $0.50~R_\odot$ case, but that the change to these structures does not significantly impact the mean MT rate. 

\section{Conclusions} \label{sec:conc}

We present here the first method to successfully account for the 3D hydrodynamics of MT in the evolution of a binary system across thermal and nuclear timescales. MT simulations on evolutionary timescales have long been limited to 1D and the substantial assumptions that requires. By hybridizing the 3D hydrodynamics code \vhone with the 1D stellar evolution code \mesa, we successfully modeled two complete MT phases in 3D, spanning the Helium-burning phase of the donor.

Our model captured both thermal- and nuclear-timescale Case B MT in a merging BBH progenitor without significant computational artifacts. We tracked the mass and AM fluxes carried by L2/L3 outflow streams without taking them into account for the evolution of the system. The binary was instead evolved under the Eddington-limited MT assumption in absence of outflows. This enabled us to track a system evolution that closely fits that of prior 1D models while obtaining the relative impact of outflow streams, were they to be considered. 

Throughout both MT phases, the L1 mass loss prescription of \cite{kolb&ritter} provided a close estimate of the results of 3D modeling. The 1D estimate following \cite{kolb&ritter} for the conditions of our 3D model remained within $20\%$ of the 3D transfer rate until the final stages of envelope stripping. Near the end of the early Case B phase and throughout the late Case B phase (g-k in Figure \ref{fig:evovkolb}), this prescription underestimated mass loss by as much as $60\%$. Our independent 1D model following \cite{kolb&ritter} predicted the final mass of the binary to within $0.002~M_\odot$. The prescription of \cite{marchant2021roleofMT} was compared directly at each stage and was found to consistently overestimate the transfer rate. The prescription of \cite{cehula2023mtprescription} was not directly computed, but as it generally returns transfer rates a factor of two lower than \cite{kolb&ritter}, we expect it would significantly underestimate the results of our 3D evolution.  

Figure \ref{fig:tconv} depicts the most significant discrepancy between our 3D model and 1D predictions. Our model remained contained by L3, suggesting secondary outflows during the end of envelope stripping may be overestimated in works that rely on 1D prescriptions (e.g. \citealp{cehula2023mtprescription}). This may result in more high mass systems remaining stable throughout MT than previously predicted.

We found dense outflow streams to form at both L2 and L3 even in systems underfilling those equipotentials. The L2 flow remained above the photospheric density for nuclear timescales, and both flows may have contributed to forming a circumbinary disk ($\lesssim0.1~M_\odot$) if modeled in a larger computational domain. The collective mass and AM loss through both streams remained negligible ($<4\%$) when compared to the total losses of the system through Eddington-limited accretion. 

Across both MT phases, the L2 stream carried $81.0\pm0.7\%$ of the specific AM of a circular orbit at the L2 point. When an L3 stream was present, it was most often consistent with carrying the specific AM of an orbit at L3. However, when L3 did vary in specific AM, fluctuations were extreme due to its low overall mass. A smooth trend occurs in the specific AM of the L1 stream, but it seems to depend significantly on the nature of the donor and/or its envelope unlike the outer streams.

Future work may more conclusively evaluate the significance of L2/L3 outflows by evolving a set of binaries in 3D across a range of mass ratios and orbital periods. Additional studies may improve our understanding of Case B MT by spanning stable Case A MT in 3D, providing clearer initial conditions to these systems. Later simulations of this kind may better examine orbital evolution and observational impacts by incorporating feedback from outflow streams and circumbinary disk formation.

\begin{acknowledgements}
We acknowledge support from KU Leuven C1 grant BRAVE C16/23/009. JO. S. acknowledges support from European Research Council (ERC) Horizon Europe under grant agreement number 101044048, and from the Belgian Research Foundation Flanders (FWO) grant G077822N. P. M. acknowledges support from the European Research Council (ERC) under the European Union’s Horizon 2020 research and innovation programme (101165213/Star-Grasp). No AI tools were employed in the creation of this work.
\end{acknowledgements}

\bibliography{references}
\bibliographystyle{bibtex/aa.bst}

\end{document}